\documentclass[twocolumn,amssymb, aps, prd, longbibliography,nofootinbib,superscriptaddress,reprint]{revtex4-1}%

\usepackage{amsmath}            
\usepackage{graphicx}
\usepackage{dcolumn}
\usepackage{bm}
\usepackage{color}
\usepackage{textcomp}
\usepackage{xcolor}
\usepackage{afterpage}
\usepackage[hidelinks]{hyperref}

\hypersetup{
  colorlinks = true, 
  urlcolor = blue!50!black, 
  linkcolor = blue!50!black, 
  citecolor = blue!50!black 
}\begin{document}
\title{Precision laser-based measurements of the single electron response of SPCs \\for the NEWS-G light dark matter search experiment}
\author{Q.~Arnaud}
\email{Corresponding author: q.arnaud@queensu.ca}
\affiliation{Department of Physics, Engineering Physics \& Astronomy, Queen's University, Kingston, Ontario K7L 3N6, Canada}
\author{J.-P.~Bard}
\affiliation{IRFU, CEA, Universit\'{e} Paris-Saclay, F-91191 Gif-sur-Yvette, France}
\author{A.~Brossard}
\affiliation{Department of Physics, Engineering Physics \& Astronomy, Queen's University, Kingston, Ontario K7L 3N6, Canada}
\affiliation{IRFU, CEA, Universit\'{e} Paris-Saclay, F-91191 Gif-sur-Yvette, France}
\author{M.~Chapellier}
\affiliation{Department of Physics, Engineering Physics \& Astronomy, Queen's University, Kingston, Ontario K7L 3N6, Canada}
\author{M.~Clark}
\altaffiliation[Now at]{ Department of Physics and Astronomy, Purdue University, 525 Northwestern Avenue, West Lafayette, IN 47907-2036}
\affiliation{Department of Physics, Engineering Physics \& Astronomy, Queen's University, Kingston, Ontario K7L 3N6, Canada}
\author{S.~Crawford}
\affiliation{Department of Physics, Engineering Physics \& Astronomy, Queen's University, Kingston, Ontario K7L 3N6, Canada}
\author{E.~C.~Corcoran}
\affiliation{Chemistry \& Chemical Engineering Department, Royal Military College of Canada, Kingston, Ontario K7K 7B4, Canada}
\author{A.~Dastgheibi-Fard}
\affiliation{LSM, CNRS/IN2P3, Universit\'{e} Grenoble-Alpes, Modane, France }
\author{K.~Dering}
\affiliation{Department of Physics, Engineering Physics \& Astronomy, Queen's University, Kingston, Ontario K7L 3N6, Canada}
\author{P.~Di Stefano}
\affiliation{Department of Physics, Engineering Physics \& Astronomy, Queen's University, Kingston, Ontario K7L 3N6, Canada}
\author{D. Durnford}
\affiliation{Department of Physics, Engineering Physics \& Astronomy, Queen's University, Kingston, Ontario K7L 3N6, Canada}
\author{G.~Gerbier}
\affiliation{Department of Physics, Engineering Physics \& Astronomy, Queen's University, Kingston, Ontario K7L 3N6, Canada}
\author{I.~Giomataris}
\address{IRFU, CEA, Universit\'{e} Paris-Saclay, F-91191 Gif-sur-Yvette, France}
\author{G.~Giroux}
\affiliation{Department of Physics, Engineering Physics \& Astronomy, Queen's University, Kingston, Ontario K7L 3N6, Canada}
\author{P.~Gorel}
\affiliation{SNOLAB, Lively, Ontario, P3Y 1N2, Canada}
\affiliation{Department of Physics and Astronomy, Laurentian University, Sudbury, Ontario, P3E 2C6, Canada}
\author{M.~Gros}
\affiliation{IRFU, CEA, Universit\'{e} Paris-Saclay, F-91191 Gif-sur-Yvette, France}
\author{P.~Gros}
\affiliation{Department of Physics, Engineering Physics \& Astronomy, Queen's University, Kingston, Ontario K7L 3N6, Canada}
\author{O.~Guillaudin}
\affiliation{LPSC, Universit\'{e} Grenoble-Alpes, CNRS/IN2P3, Grenoble, France}
\author{E.~W.~Hoppe}
\affiliation{Pacific Northwest National Laboratory, Richland, Washington 99354, USA}
\author{A.~Kamaha} 
\altaffiliation[Now at]{ Department of Physics,  University at Albany - State University of New York, Albany, New York 12205, USA}
\affiliation{Department of Physics, Engineering Physics \& Astronomy, Queen's University, Kingston, Ontario K7L 3N6, Canada}
\author{I.~Katsioulas}
\affiliation{IRFU, CEA, Universit\'{e} Paris-Saclay, F-91191 Gif-sur-Yvette, France}
\author{D.~G.~Kelly}
\affiliation{Chemistry \& Chemical Engineering Department, Royal Military
College of Canada, Kingston, Ontario K7K 7B4, Canada}
\author{P.~Knights}
\affiliation{IRFU, CEA, Universit\'{e} Paris-Saclay, F-91191 Gif-sur-Yvette, France}
\affiliation{School of Physics and Astronomy, University of Birmingham, Birmingham
B15 2TT United Kingdom}
\author{S.~Langrock}
\affiliation{Department of Physics and Astronomy, Laurentian University, Sudbury, Ontario, P3E 2C6, Canada}
\author{P.~Lautridou}
\affiliation{SUBATECH, Universit\'{e} de Nantes/Ecole des Mines de Nantes/IN2P3-CNRS, Nantes, France}
\author{R.~D. Martin}
\affiliation{Department of Physics, Engineering Physics \& Astronomy, Queen's University, Kingston, Ontario K7L 3N6, Canada}
\author{J.~McDonald}
\affiliation{Department of Physics, Engineering Physics \& Astronomy, Queen's University, Kingston, Ontario K7L 3N6, Canada}
\author{J.-F.~Muraz}
\affiliation{LPSC, Universit\'{e} Grenoble-Alpes, CNRS/IN2P3, Grenoble, France}
\author{J.-P.~Mols}
\affiliation{IRFU, CEA, Universit\'{e} Paris-Saclay, F-91191 Gif-sur-Yvette, France}
\author{K.~Nikolopoulos}
\affiliation{School of Physics and Astronomy, University of Birmingham, Birmingham
B15 2TT United Kingdom}
\author{F.~Piquemal}
\affiliation{LSM, CNRS/IN2P3, Universit\'{e} Grenoble-Alpes, Modane, France }
\author{M.-C.~Piro}
\affiliation{Department of Physics, University of Alberta, Edmonton, Alberta, T6G 2R3, Canada}
\author{D.~Santos}
\affiliation{LPSC, Universit\'{e} Grenoble-Alpes, CNRS/IN2P3, Grenoble, France}
\author{G.~Savvidis} 
\affiliation{Department of Physics, Engineering Physics \& Astronomy, Queen's University, Kingston, Ontario K7L 3N6, Canada}
\author{I.~Savvidis}
\affiliation{Aristotle University of Thessaloniki, Thessaloniki, Greece}
\author{F.~Vazquez de Sola Fernandez} 
\affiliation{Department of Physics, Engineering Physics \& Astronomy, Queen's University, Kingston, Ontario K7L 3N6, Canada}
\author{M.~Vidal} 
\affiliation{Department of Physics, Engineering Physics \& Astronomy, Queen's University, Kingston, Ontario K7L 3N6, Canada}
\author{M.~Zampaolo}
\affiliation{LSM, CNRS/IN2P3, Universit\'{e} Grenoble-Alpes, Modane, France }

\collaboration{NEWS-G Collaboration}
\noaffiliation


\begin{abstract}
Spherical Proportional Counters (SPCs) are a novel gaseous detector technology employed by the NEWS-G low-mass dark matter search experiment for their high sensitivity to single electrons from ionization. In this paper, we report on the first characterization of the single electron response of SPCs with unprecedented precision, using a UV-laser calibration system. The experimental approach and analysis methodology are presented along with various direct applications for the upcoming next phase of the experiment at SNOLAB. These include the continuous monitoring of the detector response and electron drift properties during dark matter search runs, as well as the experimental measurement of the trigger threshold efficiency. We measure a mean ionization energy of  $\mathrm{W}=27.6\pm0.2~\mathrm{eV}$ in Ne + $\mathrm{CH_4}$~(2\%) for 2.8~keV X-rays, and demonstrate the feasibility of performing similar precision measurements at sub-keV energies for future gas mixtures to be used for dark matter searches at SNOLAB.
\end{abstract}

\keywords{Single Electron Response, SPC, Polya distribution, Laser, dark matter, WIMPs}

\maketitle










\section{Introduction}
New Experiments With Spheres-Gas (NEWS-G)~\cite{NEWS-Gwebsite} is a dark matter direct detection experiment using Spherical Proportional Counters (SPCs)~\cite{SPC} to search for low-mass Weakly Interacting Massive Particles (WIMPs), a favored class of dark matter particle candidates~\cite{WIMPs}. SPCs are a novel gaseous detector technology possessing many appealing features for light dark matter searches. The high amplification gain conferred by the Townsend avalanche process allows for unprecedented sensitivity to the minute nuclear recoil energies expected from WIMPs scattering off of target nuclei. The operation of SPCs with light noble gases (He, Ne) further allows for an optimization of the momentun transfers for low-mass WIMPs, extending the expected recoil energy spectrum to higher energies. This, together with the ability to discriminate against surface events using pulse-shape analysis, has already allowed NEWS-G to set world-leading constraints for sub-GeV WIMPs with a 60 cm diameter SPC operated with $\mathrm{Ne} + \mathrm{CH_4}~(0.7\%)$ at the Laboratoire Souterrain de Modane (LSM)~\cite{NEWSGresults}. 
The next phase of the experiment, beginning in fall-2019, will see the deployment of a 140~cm SPC at SNOLAB (Sudbury, Canada) with improved shielding, higher radiopurity materials, and gas purification systems. Beyond these upgrades, the competitiveness of the experiment will be hinged upon achieving a sub-electron energy threshold and precisely characterizing the detector response at such unprecedented low energies. In pursuit of achieving the latter, we report on a powerful laser-based calibration method that will be implemented at SNOLAB. Beyond serving as a proof of concept, we present in this work the methodology and results of the first measurement of the Single Electron Response (SER) of SPCs, with percent level precision. The paper is structured as follows: in Sec.~\ref{sec:Experimental Set-Up}, we discuss the functioning principle of SPCs as well as the experimental set-up used at Queen's University (Kingston, Canada) to perform laser calibration measurements. In Sec.~\ref{sec:AnalysisMethodology}, we present the novel analysis methodology developed to obtain the mean amplification gain and its relative variance from single electron spectra. We then give in Sec.~\ref{sec:Applications} an overview of the various other useful applications for the NEWS-G experiment which extend far beyond the measurement of the SER. These include the experimental measurement of the trigger threshold efficiency as well as the monitoring of both the detector response and electron drift properties during WIMP search runs. Additionally, by combining laser-based and ${}^{37}\mathrm{Ar}$ calibration data we measure the mean ionization energy in $\mathrm{Ne} + \mathrm{CH_4}~(2\%)$ for 2.82~keV X-rays and demonstrate the feasibility of performing similar precision measurements at sub-keV energies.
\section{Experimental Set-Up}
\label{sec:Experimental Set-Up}
\subsection*{Spherical Proportional Counter (SPC)}
SPCs consist of a grounded spherical vessel acting as a cathode and a small spherical anode a few mm in diameter at its center on a support rod~(see Fig.~\ref{Fig:Methodology}). The anode is biased to a positive high voltage (HV) of up to a few thousand volts via an insulated HV wire going through the grounded support rod. The resulting electric field magnitude varies with the distance $r$ from the sensor as $1/r^{2}$, delimiting the detector volume into two regions: a large region at low-field (a few V/cm) in which electrons from ionization drift towards the sensor within typically $\sim100~\mathrm{\mu s}$, and a small amplification region in the vicinity of the anode where the high field intensity ($\sim10^5\; \mathrm{V/cm}$) triggers a Townsend avalanche. The high amplification gain of up to $10^4$ combined  with  the low intrinsic capacitance of the central sensor $\mathcal{O}(0.1~\mathrm{pF})$ results in a signal to noise ratio that provides sensitivity to single electrons from primary ionization. The drift and diffusion of electrons in the low-field region results in a measurable radial-dependent dispersion of their arrival time. This allows us to discriminate surface events associated with high rise times from bulk events for which electrons being less subject to diffusion yield pulses with smaller rise times. For the interested reader, an overview of recent developments and applications of SPCs is given in~\cite{SPC-ConfProc-RecentDev}. 
\subsection*{Laser calibration set-up}
\begin{figure*}[tpb]
\begin{center}
\includegraphics[width=0.91\textwidth]{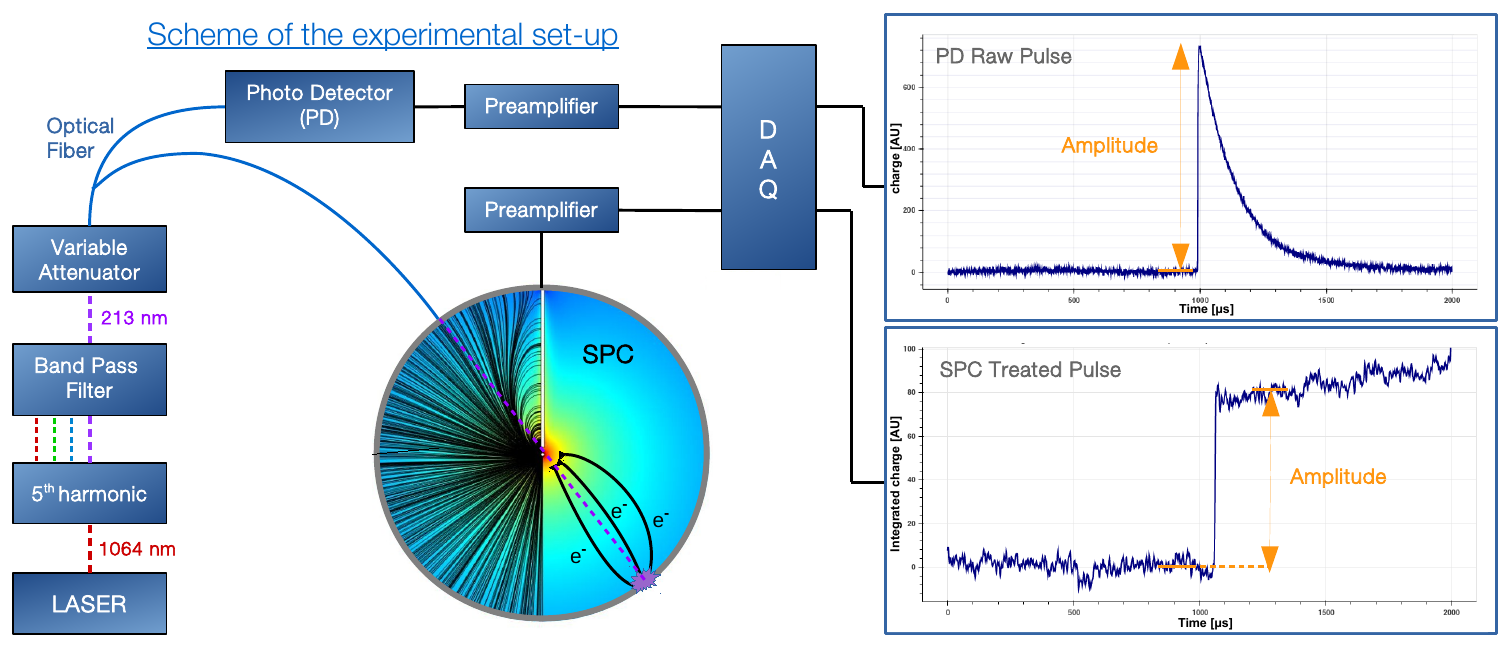}
\caption{The experimental set-up as described in Sec. \ref{sec:Experimental Set-Up}. The 213 nm laser light is sent through an optical fiber splitter to both the PD (which triggers the acquisition) and the SPC to extract photo-electrons from the inner surface of the vessel. The two panels on the right show a typical PD signal (raw pulse on top panel) together with the resulting SPC signal (treated pulse on bottom panel) from a single electron reaching the sensor and undergoing an avalanche of average gain. The time delay between the SPC and the PD pulse corresponds to the drift time of the electron from the surface to the sensor.}
\label{Fig:Methodology}
\end{center}
\end{figure*}
The inner surface of the vessel is illuminated using a monochromatic pulsed UV laser to extract a tunable number of photo-electrons down to a single electron. The experimental set-up at Queen's University to perform these laser calibration measurements is depicted in Fig.~\ref{Fig:Methodology}. 

The custom laser beam of wavelength $\lambda=1064\;\mathrm{nm}$ is produced by a compact diode-pumped solid state active Q-switched laser. This is coupled to a fifth harmonic waveform generator to produce an output beam of $\lambda=213\;\mathrm{nm}$. Left-over radiation from the first ($\lambda=1064\;\mathrm{nm}$), second ($\lambda=532\;\mathrm{nm}$), third ($\lambda=355\;\mathrm{nm}$) and fourth ($\lambda=266\;\mathrm{nm}$) harmonics are suppressed by a $213\;\mathrm{nm}$ bandpass filter. A neutral density variable attenuator with a transmission range between 1 and 100\% is used to tune the laser output to the desired power. This, along with the possiblity of adjusting the current of the Laser pump from $\sim100$~A up to 150~A, provides a dynamic range of photo-electron extraction from single electron up to a few hundred. The pulsed (10 Hz) UV laser beam is sent through an optical fiber splitter to both a Photo Detector (PD) and to the SPC. 

The PD is used to monitor the laser power, trigger the acquisition, and to determine the time at which photo-electrons are extracted. We use a Si biased PD (Thorlabs, DET10A) whose generated current of $\sim0.02\;\mathrm{A/W}$ is integrated by a charge sensitivite preamplifier (CREMAT, CCR-110-R2.1) with a 140~$\mathrm{\mu s}$ decay time constant, and the resulting voltage signal is sent to the DAQ. Fig.~\ref{Fig:Methodology} (top right) shows a typical PD raw pulse whose amplitude is proportional to the laser light power. The relative resolution of the PD signal to a fixed laser power was measured to be always of $\sim1\%$ RMS or better. This measurement was performed by replacing the SPC in Fig.~\ref{Fig:Methodology} with a second PD to disentangle fluctuations of the PD signal amplitudes arising from pulse-to-pulse instability of the laser. 

The SPC consists of a 30~cm diameter stainless steel vessel certified to hold up to $10\;\mathrm{bars}$ of gas. The 2~mm diameter anode at the center of the vessel was biased to a positive high voltage of $\mathrm{HV_{1}}\sim1000\;\mathrm{V}$ via an insulated HV wire going through a stainless steel rod of 6 mm outer diameter. The presence of the grounded rod is responsible for anisotropies of the electric field in the north hemisphere which are visible in Fig.~\ref{Fig:Methodology}. In the absence of countermeasures, these may lead to a dependence of the amplification gain on the arrival angle of primary electrons in the avalanche region~\cite{NEWSGresults,SPC-sensor-Katsioulas}. To prevent this, the sensor was supplemented with a cylindrical Bakelite electrode 14 mm in height and 14 mm in radius, placed at a distance of 6 mm from the anode. The application of a negative voltage $\mathrm{HV_2}$ on this second electrode on the order of $\mathrm{HV_{2}}/\mathrm{HV_{1}}=-10\%$ guides electrons to the bottom of the anode where the electric field is isotropic, thereby ensuring an amplification gain independent of the initial position of the event~\cite{SPC-AlexisBROSSARD}.

After pumping of the vessel down to $\sim10^{-5}~\mathrm{mbar}$, the SPC was filled with a mixture of $\mathrm{Ne}+\mathrm{CH_4}~(2\%$ in volume$)$ at 1.5 bar as well as a small amount of ${}^{37}\mathrm{Ar}$ ($\sim 15\;\mathrm{Hz}$  of events). The latter is a gaseous radioactive source producing monoenergetic events of 270 eV and 2822~eV from X-rays induced by electron capture in the L- and K-shells, respectively~\cite{Ar37properties}. The ${}^{37}\mathrm{Ar}$ source was obtained from CaO by irradiation in the predominantly thermal flux of a SLOWPOKE-2 reactor at the Royal Military College of Canada~\cite{Ar37production}. The SPC was operated in a sealed state with a recirculation system coupled to a getter to purify the gas from $O_2$ and other electronegative impurities as these result in electron attachment. Although the level of impurities was not measured directly (e.g. with a Residual Gas Analyzer), it could still be qualitatively monitored. Indeed, events with energy deposits at large radii (higher rise times) are subject to greater electron loss (smaller amplitude) in the presence of electron attachment. After a few days of gas recirculation through the getter, the suppression of attachment was confirmed by the cessation of correlations in rise time and amplitude of ${}^{37}\mathrm{Ar}$ 2822~eV events. Data acquisition was triggered by signals from the SPC and from the PD in order to record ${}^{37}\mathrm{Ar}$ events as well as all laser-induced events. Signals were sampled at $1.08\;\mathrm{MHz}$ over a 2~ms time window centered on the trigger time. As in~\cite{NEWSGresults}, SPC pulses were deconvolved for the detector response (i.e. ion induced current convolved with the pre-amplifier response). Treated pulses consist of the cumulative integral of the deconvolved pulses, which to first-order are series of individual step functions centered on the arrival time of primary electrons, and whose amplitudes are proportional to the number of secondary electron/ion pairs produced in the avalanche. 
As an illustrative example, we show in Fig.~\ref{Fig:Methodology} the treated pulse of a typical laser-induced single electron event, delayed with respect to the PD pulse by the drift time of the electron. The amplitudes of laser-induced events were evaluated at a fixed time of $\sim200~\mathrm{\mu s}$ after triggering on the PD channel to ensure all photo-electrons had time to reach the sensor, irrespective of the run voltage conditions. Fixing the pulse end-time is a necessary component of the analysis methodology described in the following section, since it allows for modeling the effect of baseline noise fluctuations on the amplitude of all laser-induced events with a Gaussian distribution.
\section{Analysis methodology}
\label{sec:AnalysisMethodology}
\subsection{Modeling of the detector response}
\label{sec:model1}
The Single Electron Response (SER) of the SPC is driven by the statistics of secondary ionization of the Townsend avalanche. The number of secondary electron/ion pairs produced is stochastic and subject to large statistical fluctuations. Under the condition that the ionizing probability of an electron within an avalanche is independent of its past history, the distribution of the number of ions produced is a decaying exponential according to Fury law~\cite{sauli_fury}. The fulfillment of this condition depends on various parameters including detector geometry, gas characteristics, electric field $E$ to pressure $P$ ratio $E/P$ and the gain. Experimental measurements in Micromegas~\cite{MICROMEGAS-POLYA-DERRE,MICROMEGAS-SER} and GEMs~\cite{GEM-SER,GEM-IMAGING-Polya} show that at high $E/P$ values and high gain, the SER is better decribed by the so-called Polya distribution:
\begin{eqnarray}
P_{\mathrm{Polya}}(S)&=&\dfrac{1}{\langle G \rangle }\frac{\cdot (1+\theta)^{1+\theta}}{\Gamma(1+\theta)}\left({\frac{S}{\langle G \rangle }}\right)^{\theta} \nonumber \\ 
&& \times \; 
\; \mathrm{exp}\left({-(1+\theta)\frac{S}{\langle G \rangle } }\right)
\label{Polya}
\end{eqnarray}
where $S$ is the number of secondary electron/ion pairs produced in the avalanche and $\langle G \rangle$ is the mean gain. The parameter $\theta $ drives the shape of the probability distribution function from an exponential ($\theta=0$) which corresponds to the Fury distribution, to a normal distribution ($\theta\gg1$). It also determines the dispersion of the avalanche gain fluctuations whose relative variance, often denoted as $f$, is given by:
\begin{equation}
f=\frac{1}{1+\theta}
\label{eq:f}
\end{equation}
Considering electron avalanches as being independent from one another, the probability of creating $S$ secondary electron/ion pairs when $N$ primary electrons reach the avalanche region is given by the $N^{\mathrm{th}}$ convolution of the Polya distribution: 
\begin{eqnarray}
P_{\mathrm{Polya}}(S|N)&=&\frac{1}{\langle G \rangle} \left(\frac{(1+\theta)^{1+\theta}}{\Gamma(1+\theta)}\right)^N\left({\frac{S}{\langle G \rangle }}\right)^{N(1+\theta)-1} \nonumber \\
& & \times \; \; \mathrm{exp}\left({-(1+\theta)\left({\frac{S}{\langle G \rangle }}\right)}\right) \nonumber \\
& & \times \; \;\prod_{j=1}^{N-1} B\left((j+j\theta),(1+\theta) \right)
\end{eqnarray}
where $B(x,y)$ is the beta function. By definition, when N=1 we recover Eq.~\eqref{Polya}. 

The photo-electric effect from $\lambda=213\;\mathrm{nm}$ radiation is linear such that the mean number $\mu$ of primary electrons extracted from the inner surface of the vessel is expected to be proportional to the laser light intensity, and hence to the PD amplitude. For a fixed laser intensity, the actual number of primary electrons extracted $N_{\mathrm{ext}}$  is subject to statistical fluctuations anticipated to follow the Poisson distribution:
\begin{equation}
P_{\mathrm{Poisson}}(N_{\mathrm{ext}}|\mu)=\frac{ e^{-\mu}{\mu}^{N_{\mathrm{ext}}} }{N_{\mathrm{ext}}!}
\label{eq:Poisson}
\end{equation}
Because the acquisition is triggered with the PD, measured energy spectra contain so-called ``null-events'' for which no photo-electron was extracted, so that the recorded trace on the SPC channel is devoid of signal. For these events, the measured amplitude is effectively zero to within baseline noise fluctuations such that the detector response can be described by a gaussian centered on zero with standard deviation $\sigma$. For events with $N>0$ electrons, we model the detector response with the $N^{\mathrm{th}}$ convolution of the Polya distribution convolved with a Gaussian with the same $\sigma$ to account for the effect of baseline noise fluctuations on the determination of the amplitude of the events. From the above, the probability distribution function $P(E)$ of energy $E$ for laser-induced events can be written as:
\begin{equation}
P(E)=\frac{1}{\sqrt{2\pi}\sigma}\int_{-\infty}^{\infty}f(E') \cdot e^{-\dfrac{(E-E')^2}{2\sigma^2}} \mathrm{d}E'
\label{eq:pdfFit}
\end{equation}
where
\begin{eqnarray*}
f(E')=P_{\mathrm{Poisson}}(0|\mu)
 + \sum_{N=1}^{\infty}P_{\mathrm{Polya}}(E'|N) \cdot P_{\mathrm{Poisson}}(N|\mu) 
\end{eqnarray*}
A binned maximum likelihood fit of the model to the energy spectrum is performed using the following log-likelihood function:
\begin{equation}
\mathcal{L}(\langle G \rangle,\theta,\sigma,\mu)= -\mu \sum_{i=1}^{N_{\mathrm{bins}}} N_i \cdot \mathrm{log}\left(N_{\mathrm{tot}}\int_{\Delta_i}^{}P(E')dE'\right)
\label{eq:singlefit}
\end{equation}
where $N_i$ refers to the number of events observed in the $i^{\mathrm{th}}$ bin of width $\Delta_i$, while $N_{\mathrm{tot}}$ refers to the total number of events which is fixed by the acquisition time and laser frequency (i.e. independent of the energy range over which the data is fit). The mean gain $\langle G \rangle$, Polya distribution parameter $\theta$, mean number of primary electrons extracted $\mu$ and standard deviation of the baseline noise $\sigma$ are left as free parameters in the fit.

The model described here assumes a fixed value of $\mu$, and thereby of the laser power. Because of the pulse to pulse instability of the laser ($\sim 20\%$ relative dispersion), each data set was divided into sub-sets of fixed PD amplitudes to within $\pm\;5\%$. In addition to fitting each sub-set independently, a joint likelihood fit of the sub-sets is performed by maximizing the following function:
\begin{equation}
\mathcal{L}(\langle G \rangle,\theta,\sigma,\vec{\mu})= \sum_{j=1}^{N_{\mathrm{data}}}\mathcal{L}(\langle G \rangle,\theta,\sigma,\mu_j)
\label{eq:multifit}
\end{equation}
where the notation $\vec{\mu}=\{\mu_1,\mu_2,...,\mu_{N_{\mathrm{data}}}\}$ is used to denote the fact that the value of $\mu$ in each of the $N_{\mathrm{data}}$ data sets is different and left as a free parameter. This approach utilizes the whole data set, thereby ensuring the most efficient extraction of the information from each run. It also allows for the analysis of different runs taken in the same conditions but at varying laser light intensities. 
\subsection{Analysis approach}
\label{sec:Analysisapproach}
\begin{figure*}[tpb]
\begin{center}
\includegraphics[width=0.495\textwidth]{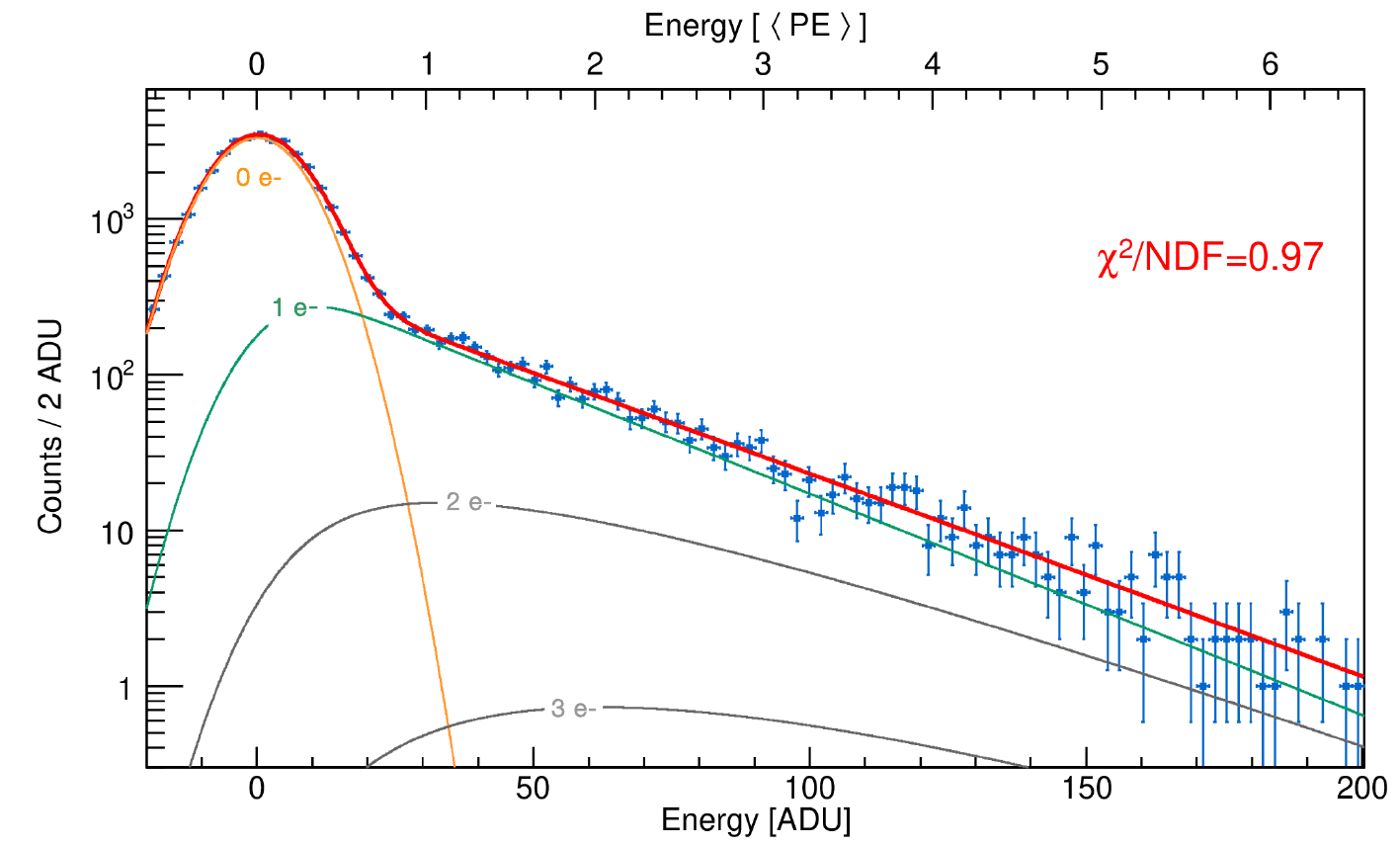}
\includegraphics[width=0.495\textwidth]{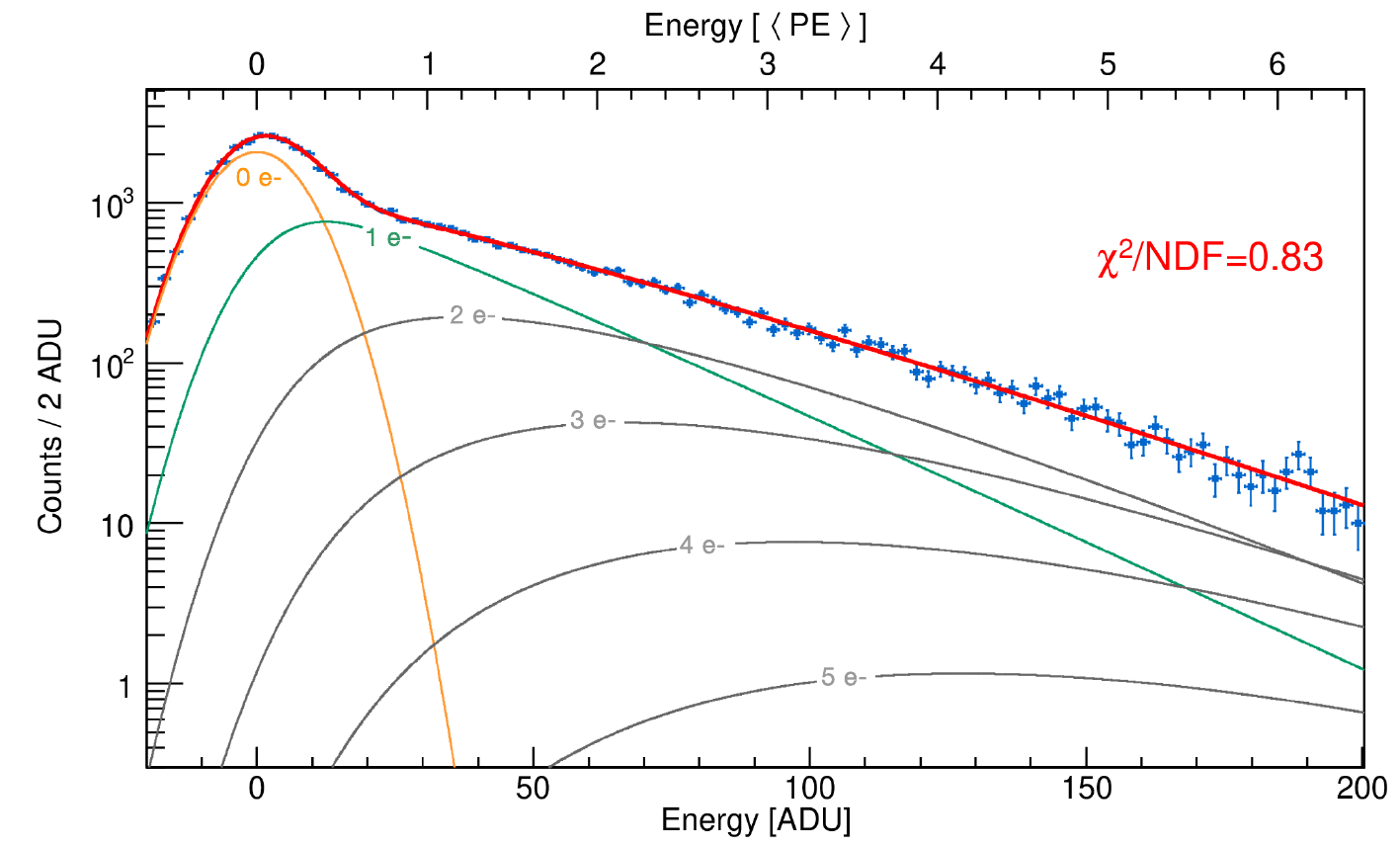}
\caption{Energy spectra (SPC channel) of laser-induced events associated with low (left panel) and high (right panel) PD pulse amplitudes. In both panels, the fit of our model (described in Sec.~\ref{sec:model1}) to the data is shown as a solid red line. The relative contribution of the null-events, single and multiple electron events are shown as solid orange, green and grey lines, respectively. The top axis gives the energy scale in average number of primary electrons $\langle\mathrm{PE}\rangle$ based on the best fit value of the mean gain. The reduced $\chi^{2}$ of the fit (normalized to the 121 degrees of freedom) is indicated as well. }
\label{Fig:Fit}
\end{center}
\end{figure*}
\begin{figure*}[tpb]
\begin{center}
\includegraphics[width=0.495\textwidth]{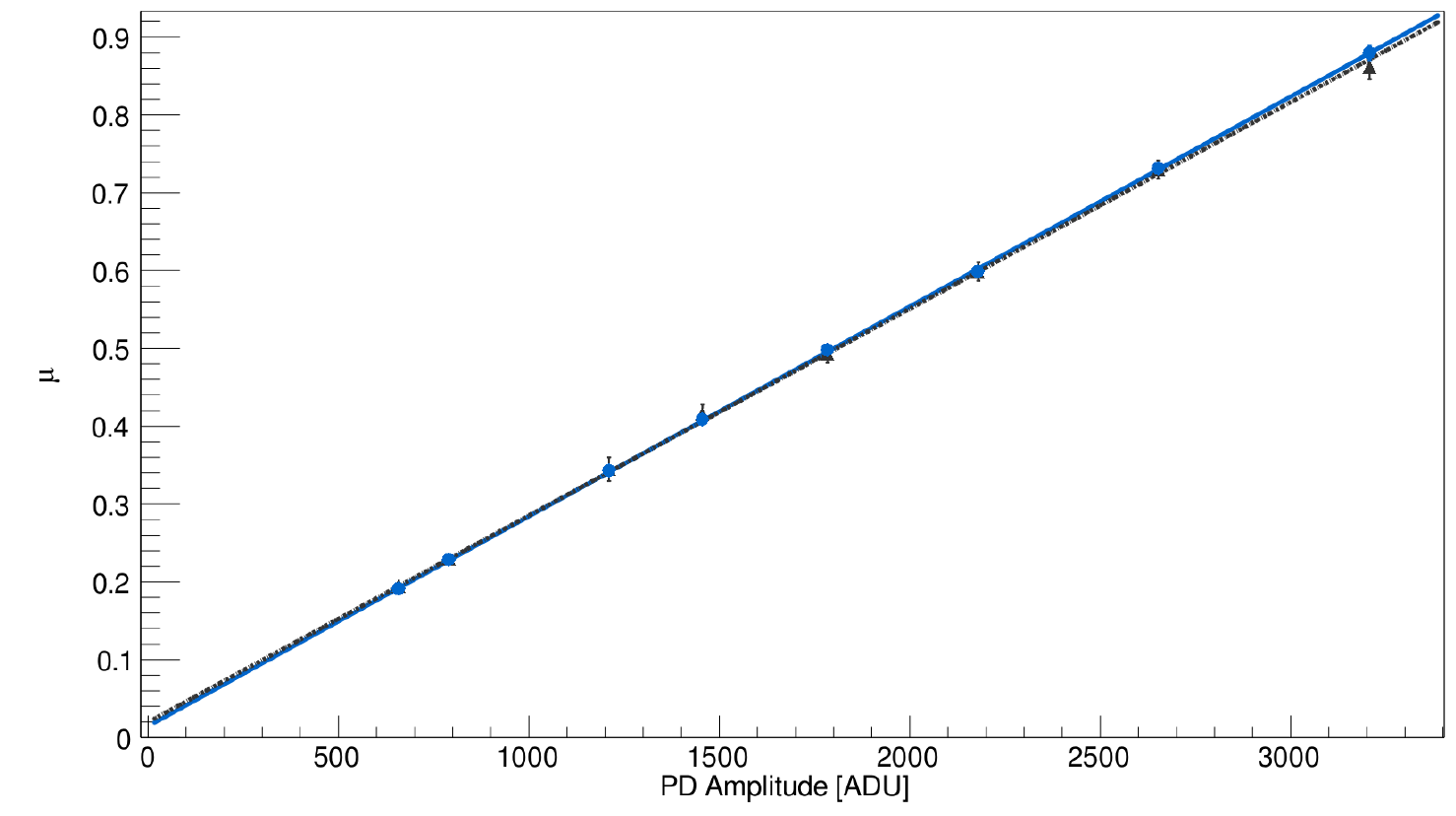}
\includegraphics[width=0.495\textwidth]{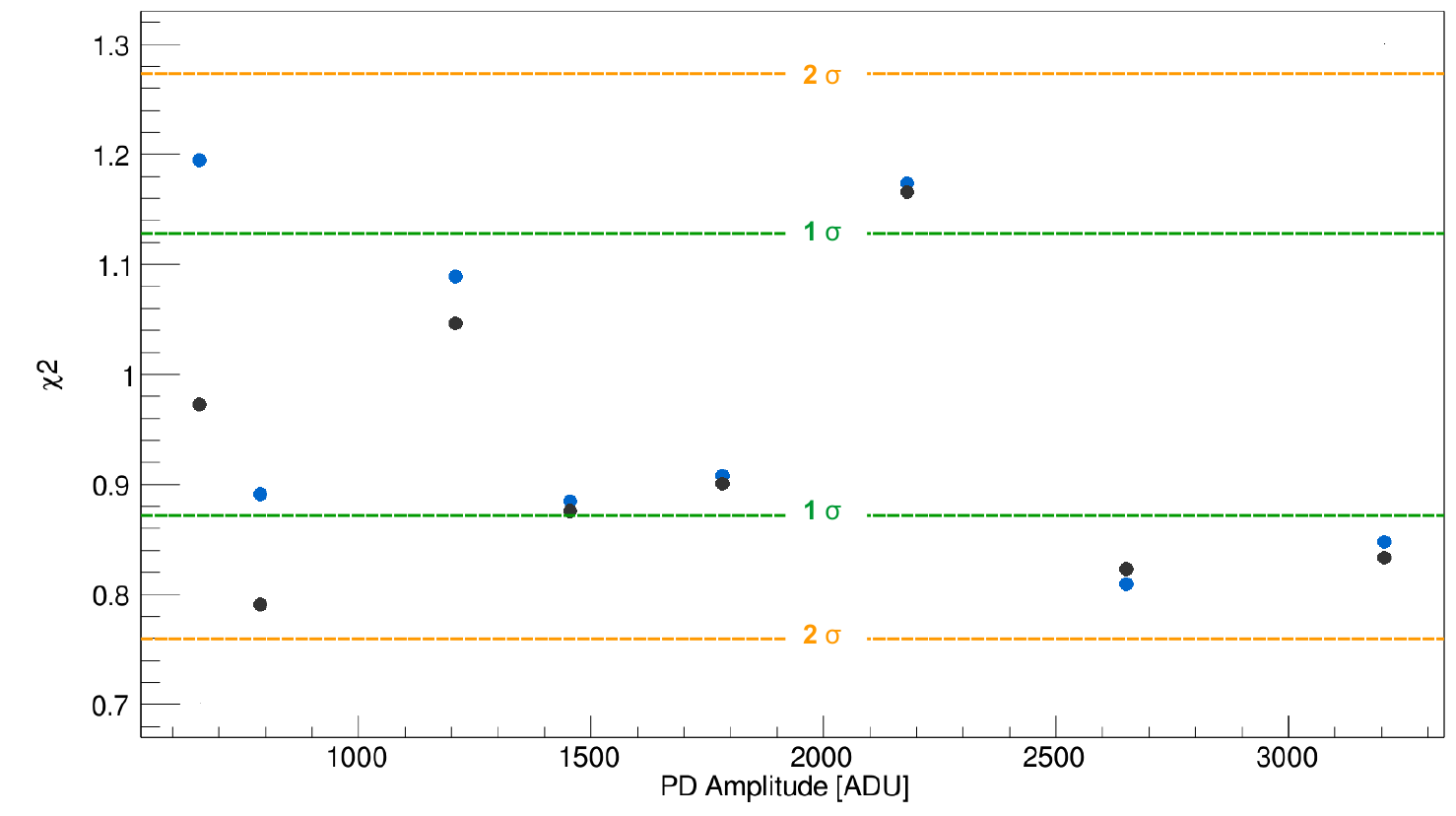}\\
\includegraphics[width=0.495\textwidth]{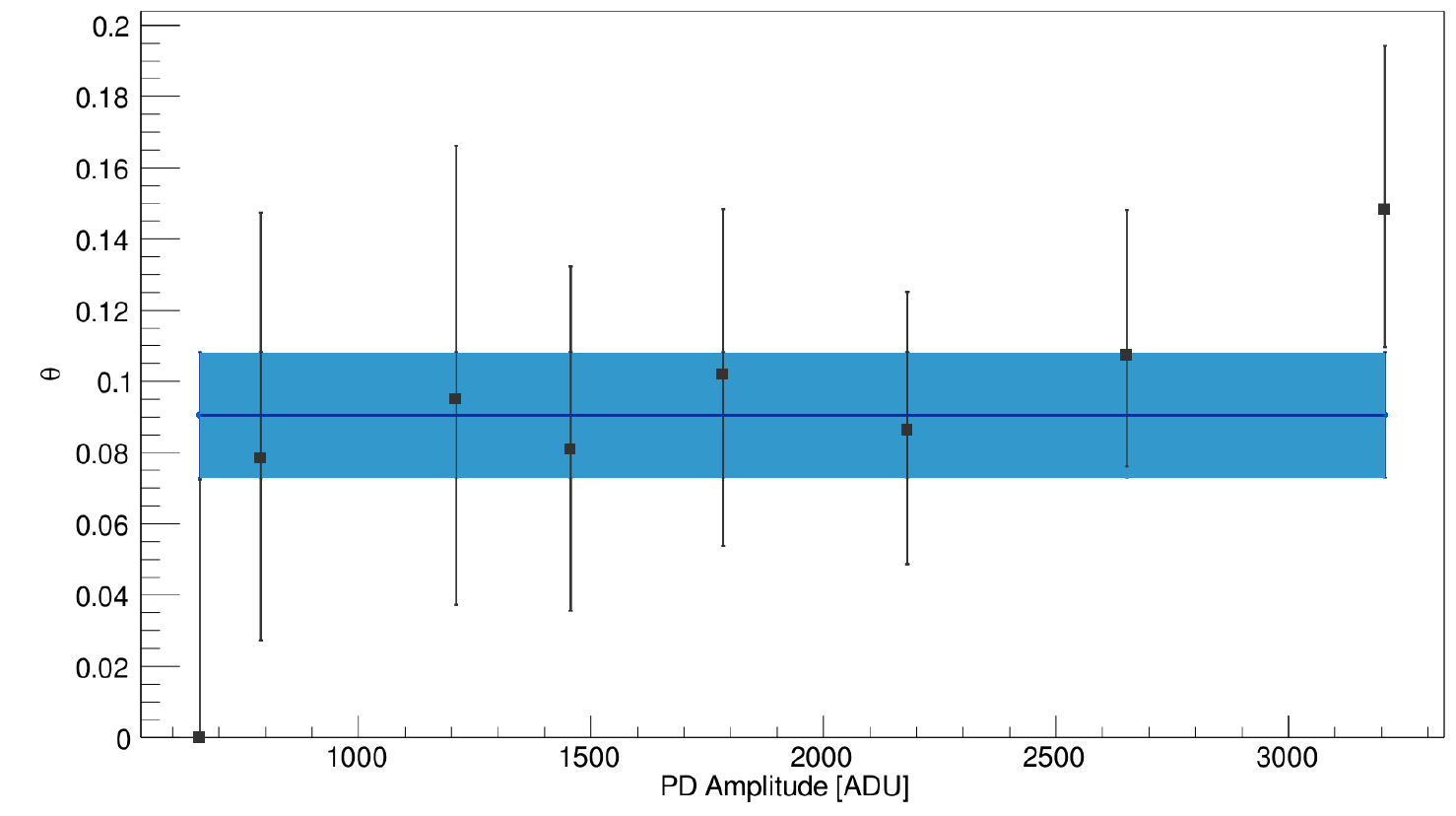}
\includegraphics[width=0.495\textwidth]{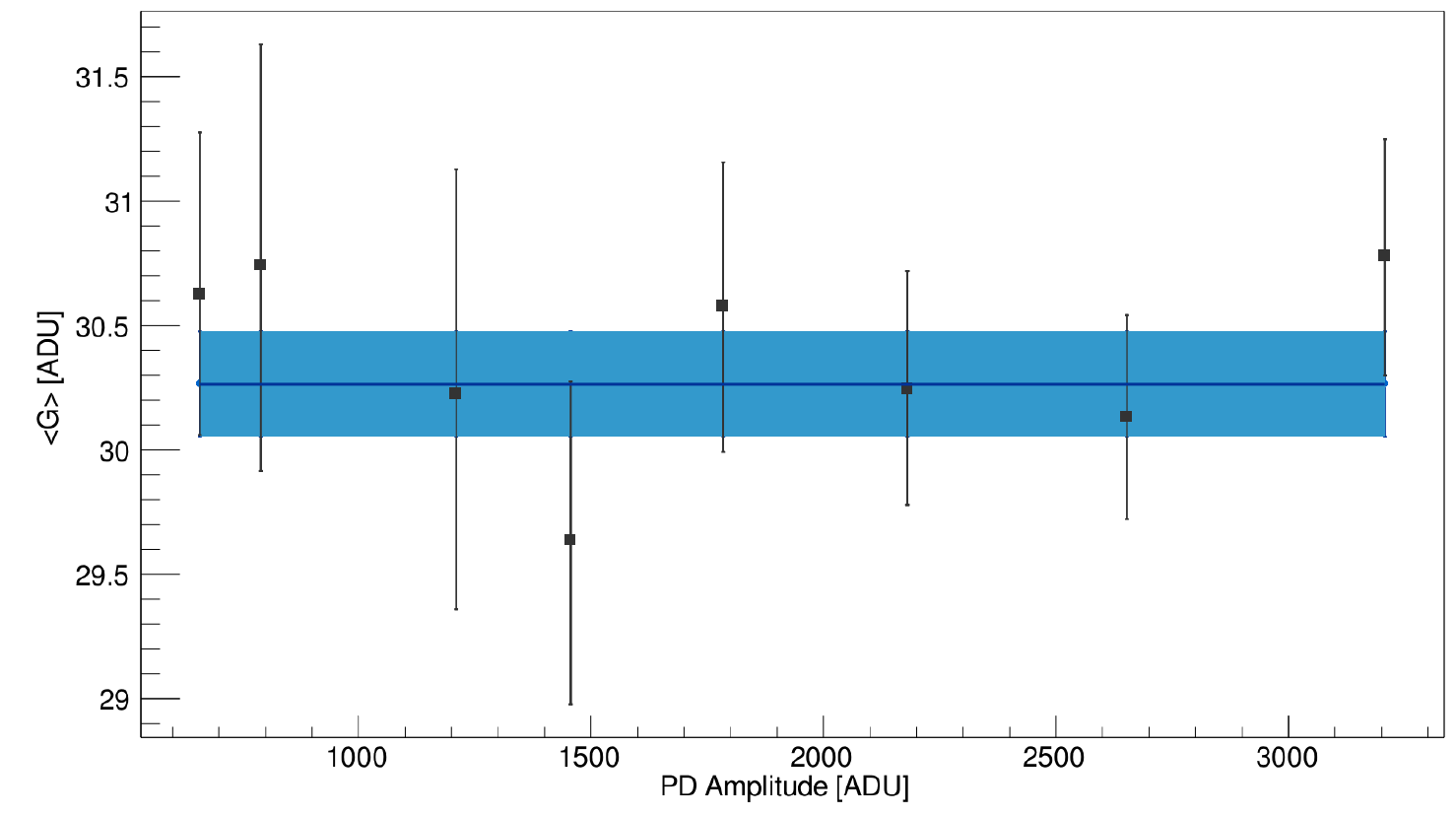}
\caption{Results from the fit of the energy spectra for laser-induced events recorded during the same run as in Fig.~\ref{Fig:Fit}. On all panels results are reported as a function of the mean amplitude of the PD signals in each of the 8 data sets. The color code indicates if the results are obtained from performing an individual fit (black) or a joint fit of all sub-sets (blue). 
Top left panel: markers correspond to the best fit values of the mean number of electrons $\mu$ while the two lines show the result from the fit of a first-order polynomial to the markers of the matching color. Both methods confirm the linearity between $\mu$ and the laser power. Top right panel: values of the $\chi^2$ normalized to the number of degrees of freedom  for each sub-set are shown both for the individual fits (black dots) and joint fit (blue dots). The green (resp. orange) dashed lines correspond to the $1~\sigma$ (resp. $2~\sigma$) confidence intervals derived from the chi-squared distribution with 121 degrees of freedom to assess the goodness of individual fits. Bottom panels: black markers correspond to the best fit values of $\theta$ (left) and of the mean gain $\langle G \rangle$ (right) for each data set. The best fit values from the joint fit are reported as a solid blue line. The light blue band corresponds to the $1~\sigma$ confidence region calculated with a profile likelihood approach.  These correpond to $ \theta = 0.09 \pm 0.02 $ and $ \langle G \rangle = 30.26 \pm 0.21~\mathrm{ADU}$. Note that $1~\mathrm{ADU}=186\pm19~\mathrm{SE}$ (secondary electrons).
}
\label{Fig:Fit2}
\end{center}
\end{figure*}
Measurements of the SER have been performed for amplification gains in the range of $10^3$ to $10^4$ by varying the voltage $\mathrm{HV_1}$ applied on the sensor from 1100~V to 1250~V. $\mathrm{HV_2}$ was always set such that $\mathrm{HV_2}/\mathrm{HV_1}=-10\%$ in order to ensure that the electric field geometry is independent of the run conditions. Each data-set consists of a run of 4 to 16 hours with the laser transmission being varied every couple of hours.

The position of the ${}^{37}\mathrm{Ar}$ 2822~eV peak was used to monitor the stability of the gain over time, allowing us to reject a few runs for which fluctuations were larger than $2\%$. To ensure the SER spectra are free from non laser-induced events, only events for which the PD channel triggered were included in the analysis. Additionally, a cut was applied requiring either the trigger time of the SPC channel to fall within $200\;\mathrm{\mu s}$ after the trigger time of the PD channel or an absence of trigger on the SPC channel (to keep null-events). The selection efficiency of this cut is energy independent. 
To illustrate the analysis procedure in more details, we use as an example one of the runs taken with $\mathrm{HV_1}$=1200~V which was divided into 8 subsets based on the amplitude of the PD signals. These data sets were fit both independently (Eq.\eqref{eq:singlefit}) and jointly (Eq.\eqref{eq:multifit}). 

We show in the left panel (resp. right panel) of Fig.~\ref{Fig:Fit} the energy spectra of the SPC signals measured for the sub-data sets with the lowest (resp. highest) laser power, together with a fit of our model to the data (solid red line). The relative contribution of the null-events (orange line), single electron events (green line), and multiple electron events (grey lines) are fixed by the Poisson distribution with mean $\mu$. In both panels, the energy scale is given in ADU and in average number of primary electrons $\langle \mathrm{PE} \rangle$ based on the best fit value of the mean gain $\langle G \rangle$. Note that $1~\mathrm{ADU}=186\pm19~\mathrm{SE}$ (secondary electrons). As attested to by the goodness of fit, our model is in excellent agreement with the data. Additionally, the analysis approach allows for the SER to be determined even in the presence of a large proportion of events with $N>1$ electrons. 

To demonstrate the overall consistency of the agreement between data and model, in Fig.~\ref{Fig:Fit2} we show as a function of the mean PD amplitude of each sub-data set, the $\chi^{2}$ as well as the best fit values of our parameters of interest ($\mu$, $\theta$, and $\left<G\right>$) obtained from both the independent (black markers) and joint (blue markers and bands) fits of each of the 8 sub-datasets. All reduced $\chi^{2}$ are close to unity and are naturally slightly lower for the independent fits as these are less constrained. As expected from our model and the linearity of the photo-electric effect at 213 nm, we do observe a linear dependance of $\mu$ with the PD amplitude. The best fit values of $\theta$ and $\left<G\right>$ for each individual data set are as we expect; they are independent of the PD amplitude to within the error bars, and consistent with the results obtained from the joint likelihood fit. 

The experimental approach together with this novel analysis methodology allows for a precision characterization of the SER, with $\mathcal{O}(1\%)$ precision measurements of the mean gain and of its relative variance $f$. The uncertainties reported in Fig.~\ref{Fig:Fit} are statistical only and determined using a profile likelihood approach. However, this approach yields measurements of $\theta$ and of the mean gain $\langle G \rangle$ that are extremely robust against mis-modeling arising from laser power fluctuations and are unaffected by electron attachment (see Appendix). 

The impact of other possible systematic uncertainties were assessed by studying the reproducibility of the above measurement through the analysis of multiple runs taken within the same voltage conditions over the course of a few weeks. Because data taking was not performed in a temperature-controlled room, the gain was subject to long term variations (see Sec.~\ref{sec:monitoring}) larger than the precision of our measurements. Therefore gain measurements from the fit of SER spectra were compared to the relative position of the ${}^{37}\mathrm{Ar}$ 2822~eV peak. A direct by-product of this study is a measurement of the mean ionization energy, which is discussed in the following section.
\section{Applications}
\label{sec:Applications}
\subsection{SER and W-value measurements }
\label{sec:W-value}
The NEWS-G experiment's sensitivity to the lowest WIMP masses may primarily derive from the detection of single electrons. Therefore, a precision characterization of the SER is critical. Additionally, because the ionization properties of noble gases mixtures such as $\mathrm{Ne}$ or $\mathrm{He}$ with $\mathrm{CH_4}$ have not been widely studied, mean ionization energy (W-value) measurements are also crutial. 

In this section, we report the results from laser-based calibration measurements in $\mathrm{Ne} + \mathrm{CH_4}~(2\%)$ gas at 1.5~bar using the approach presented in Sec~\ref{sec:Analysisapproach}. Series of runs were first taken with the same operating condition ($\mathrm{HV_1}=1200~\mathrm{V}$) to check the reproducibility of the measurements. From these, we derive the first precision characterization of the SER in SPCs, which is found to be well described by the Polya distribution with $\theta=0.12\pm0.03$. Measurements performed at different amplification gains in the range of $\sim 10^3$ ($\mathrm{HV_1}=1100~\mathrm{V}$) to $\sim 10^4$ ($\mathrm{HV_1}=1250~\mathrm{V}$) yielded similar results to within uncertainties. Because various other parameters such as the gas mixture and sensor geometry may affect the SER, laser-based calibration measurements will have to be performed in situ for the NEWS-G experiment at SNOLAB. Still, the above results demonstrate the feasability of $\mathcal{O}(1\%)$ precision measurements of $\theta$, which is more than adequate for WIMP sensitivity calculation to be robust against SER mis-modeling. 

The high statistics of ${}^{37}\mathrm{Ar}$ events (see Fig.~\ref{Fig:2820eV} in Sec~\ref{sec:understanding}) allowed for a precision determination of the mean 2822~eV peak position in each individual laser calibration run. Combining this information with measurements of the mean amplification gain $\langle G \rangle$, one can derive a measurement of the mean ionization energy W for X-rays. 
W-value measurements were also performed in various voltage conditions and were all found to be consistent to within uncertainties. 

From these, we derive a measurement of $\mathrm{W}=27.6\pm0.2~\mathrm{eV}$ in $\mathrm{Ne} + \mathrm{CH_4}~(2\%)$ at 1.5 bar for 2822~eV X-rays. Although we found no existing W-value reported for $\mathrm{Ne}$-$\mathrm{CH_4}$ gas mixtures to compare with our result, it should be mentioned that this value is significantly lower than existing measurements in pure Ne ($\mathrm{W}\sim 36~\mathrm{eV}$~\cite{WvalueNeALL,WvalueNe2}). It is also worth emphasizing that even though the getter ensured the suppression of electron attachment, this could only have led to an overestimation of the W-value (see Appendix) and therefore cannot be the cause of such a low value. In pure methane, we expect the W-value for $2822~\mathrm{eV}$ X-rays to be $\mathrm{W}\sim 27.7~\mathrm{eV}$~\cite{WAIBEL,WvalueMethane}. The similarity between the W-value we measure in $\mathrm{Ne} + \mathrm{CH_4}~(2\%)$ with that of the value for pure methane is evidence of a strong contribution of Penning effects on the ionization yield, processes by which for example, $\mathrm{Ne}^{*}$ atoms with an excitation energy higher than the ionization potential of $\mathrm{CH_4}$ can ionize the latter~\cite{PenningDefinition}. Although the magnitude of the impact on the W-value may seem surprising given the low concentration of methane, such an effect has already been reported in Ne-Xe gas mixtures~\cite{PenningWvalueNeXe}.
\subsection{Understanding of the energy resolution}
\label{sec:understanding}
In SPCs, and more generally in proportional counters, the relative energy resolution achievable for a monoenergetic peak is limited by the avalanche gain fluctuations and primary ionization statistics. It is related to the number of primary electrons by~\cite{MICROMEGAS-SER}:
\begin{equation}
\left(\dfrac{\sigma(E)}{E}\right)^2 = \frac{f}{\mu}+\dfrac{F}{\mu} + \left(\dfrac{\sigma_{b}}{\left<G\right>\mu}\right)^2 
\label{eq:reso1} 
\end{equation}
where $\sigma_{b}$ is the standard deviation of the baseline noise, $\left<G\right>$ the mean gain, $f$ the relative variance of the gain and $F$ the Fano factor defined as the ratio $\sigma_N^2/\mu$ of the variance to the mean of the number of primary electrons created $N$. Because $\left<G\right> \gg \sigma_{b}$, the last term of Eq.~\eqref{eq:reso1} is of second order and becomes even completely negligible when $\mu\gg 1$. Based on this and using Eq.~\eqref{eq:f}, we reexpress Eq.~\eqref{eq:reso1} as a function of the energy deposited in the gas $E_d$ and of the mean ionization energy $W(E_d)$ as follows:
\begin{equation}
\left(\dfrac{\sigma(E)}{E}\right)^2 =\frac{W(E_d)}{E_d} \left(\frac{1}{1+\theta}+ F(E_d) \right)
\label{eq:resoultime}
\end{equation}
Eq.~\eqref{eq:resoultime} becomes particularly interesting when $\theta$ and $W(E_d)$ are known, as one can then derive from the relative energy resolution to a monoenergetic line a measurement of the Fano factor. Additionally, in spite of the strong assymetry of the Polya distribution, the $N^{\mathrm{th}}$ convolution of the Polya distribution converges - as one expects from the central limit theorem - to a normal distribution when $N\gg 1$. In a such case, the energy resolution $\sigma$ can be measured from the simple fit of a Gaussian.
\begin{figure}[btp]
\begin{center}
\includegraphics[width=0.48\textwidth]{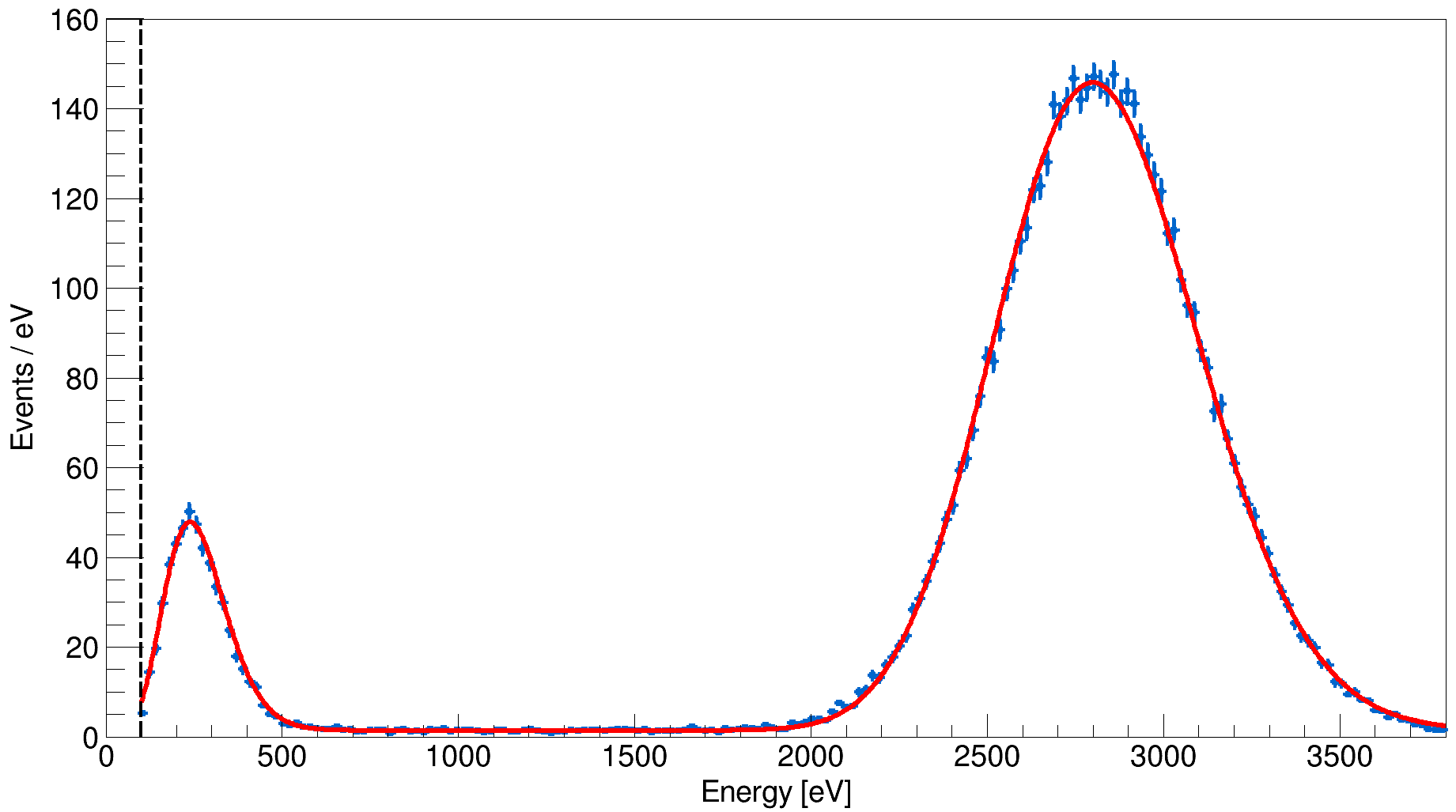}
\caption{Energy spectrum of non laser-induced events recorded during laser calibration measurements in $\mathrm{Ne} + \mathrm{CH_4}~(2\%)$ at 1.5 bar with $\mathrm{HV_1}= 1150~\mathrm{V}$. The spectrum clearly shows the 270 eV and 2822 eV lines of X-rays from electron capture in the L- and K-shell of ${}^{37}\mathrm{Ar}$, respectively. The energy scale is determined based on the position of the 2822 eV peak. The dashed line indicates the analysis threshold that was set at 100~eV. The solid red line indicates the fit of our model to the data. Our modeling of the detector response accounts both for primary ionization statistics with the COM-Poisson distribution and for statistical fluctuations of the avalanche gain with the Polya distribution. See core text for more details.}
\label{Fig:2820eV}
\end{center}
\end{figure}
We show in Fig.~\ref{Fig:2820eV} the ${}^{37}\mathrm{Ar}$ energy spectrum recorded during one of the laser calibration measurements with $\mathrm{HV_1}= 1150~\mathrm{V}$. Only non laser-induced events were selected and cuts in rise time were applied to maximize the purity in ${}^{37}\mathrm{Ar}$ events. 
The solid red line shows a fit of the 270~eV and 2822~eV peaks together with a flat background component. The spectrum was fitted only down to 100~eV to ensure the signal efficiency of the cuts in rise time is energy independent on the analysis range. Although the 2822 eV line could be fitted with a Gaussian ($N\sim100$), the 270 eV line could not ($N\sim10$), which is why we used the following probability distribution function:
\begin{equation}
\mathcal{P}(E)=\sum_{N=1}^{\infty}P_{\mathrm{Polya}}(E|N) \cdot P_{\mathrm{COM}}(N|\mu(E_d),F(E_d))
\label{eq:fitAr}
\end{equation}
where $P_{\mathrm{com}}(N|\mu(E_d),F(E_d))$ is derived from the COM-Poisson distribution~\cite{COM-Poisson-Revival,COM-Poisson-FirstPaper}, a discrete distribution function well-suited to model ionization statistics as it allows for an independent control - and hence fitting - of the mean value $\mu(E_d)=E_d/W(E_d)$ and Fano factor $F(E_d)$~\cite{DANcompoisson}. The energy scale of the ${}^{37}\mathrm{Ar}$ spectrum shown in Fig.~\ref{Fig:2820eV} was determined based on the position of the 2822 eV K-line. A binned likelihood fit of the energy spectrum was performed from modeling the detector response to mono-energetic radiation with Eq.~\ref{eq:fitAr}. The relative abundance of 270 eV and 2822 eV events from X-rays induced by electron capture in the L- and K- shells was fixed to the expected branching ratio L/K of $0.0987$~\cite{Ar37properties}. The values $\theta=0.12$ and $\mathrm{W_K}=27.6~\mathrm{eV}$ were fixed according to our measurements (see Sec.~\ref{sec:W-value}). This yields best fit values of the mean ionization energy $\mathrm{W_L}=27.6~\mathrm{eV}$, and of the Fano factor $\mathrm{F_L}=0.26$, and $\mathrm{F_K}=0.19$. Because our modeling of the detector response may not account for all possible sources of degradation of the resolution, Fano Factor best fit values should be considered as upper limits only. To assess the extent with which the above values depend on our modeling of the SER, the energy spectrum was also fitted by fixing $\theta$ to extreme values based on the precision of our measurement of $\theta=0.12\pm0.03$. Choosing $\theta=0.09$ ($\theta=0.15$) instead yields best fit values of $\mathrm{W_L}=27.8~\mathrm{eV}$ ($\mathrm{W_L}=27.4~\mathrm{eV}$), $\mathrm{F_L}=0.21$ ($\mathrm{F_L}=0.30$) and $\mathrm{F_K}=0.16$ ($\mathrm{F_K}=0.21$). 

Because we did not investigate the possible sources of systematics associated with these measurements as thoroughly as we did for $\mathrm{W_K}$ and $\theta$, the above results should not be misinterpreted as reported measurements. These are intended to demonstrate the \textit{feasability} of performing precision measurements of the mean ionization energy at sub-keV energies and of setting constraints on the Fano Factor. Furthermore, these are evidence of the excellent agreement between our model and the energy response of the detector. It is also worth emphasizing that this study allowed for an assessment of the COM-Poisson distribution as a model for primary ionization statistics~\cite{DANcompoisson}.
\subsection{Trigger Efficiency}
An accurate calculation of the detection efficiency of low-energy events requires a precise determination of the trigger threshold efficiency. It is even more critical in the context of light dark matter searches, where an experiment's sensitivity to the lowest WIMP masses may entirely depend upon events with sub-threshold energy depositions only detectable due to upwards fluctuations of the baseline noise. For the first results of the NEWS-G experiment at the LSM, the trigger threshold efficiency was obtained by applying the on-line trigger algorithm to simulated pulses added on top of realistic baseline noise samples taken from the pre-traces of real pulses recorded during the WIMP search run~\cite{NEWSGresults}. 

\begin{figure}[btp]
\begin{center}
\includegraphics[width=0.52\textwidth]{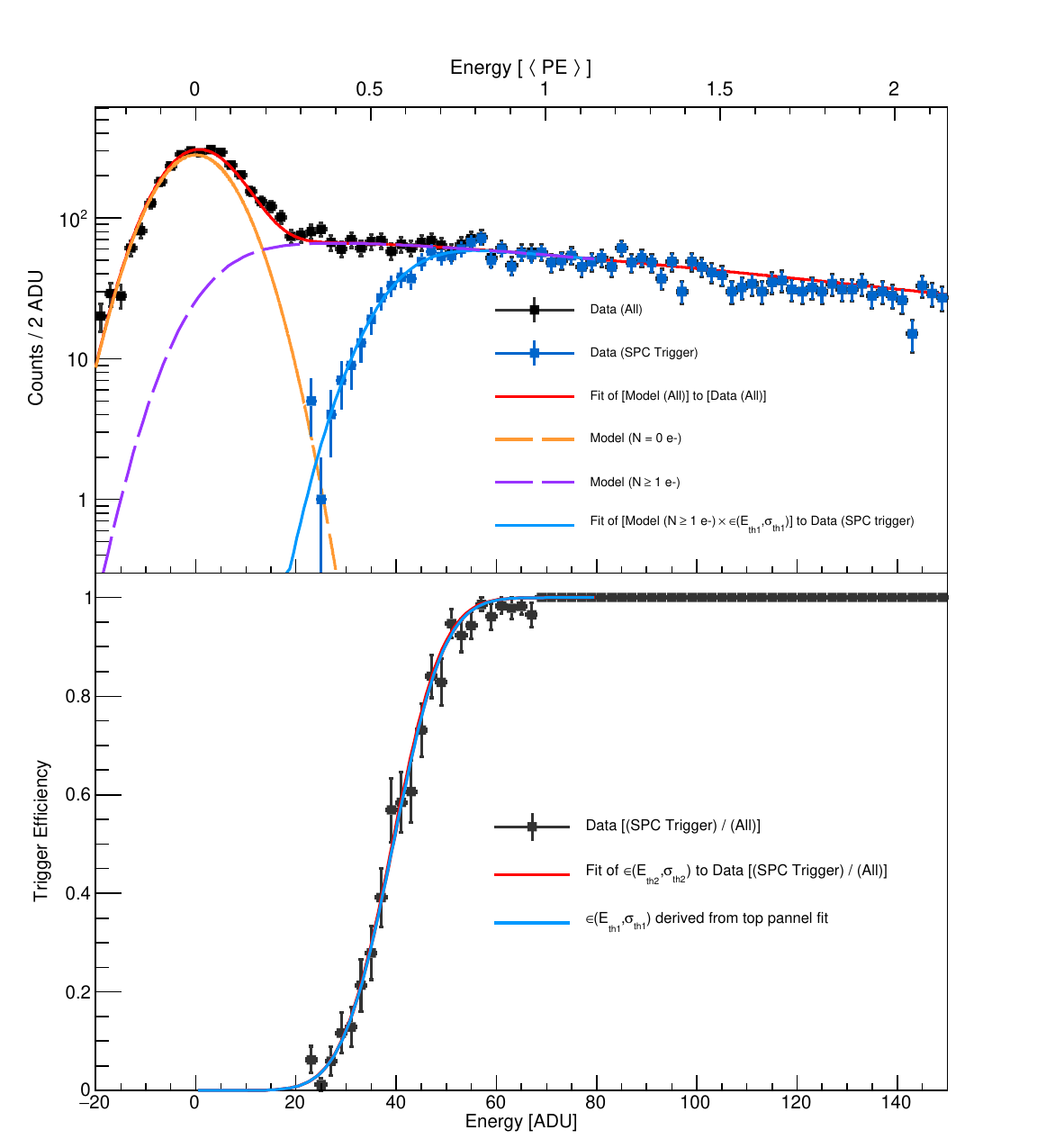}
\caption{Top Panel: Energy spectrum of laser-induced events with (blue markers) or without (black markers) triggering on the SPC channel. The energy scale is indicated - on the top axis - in average number of primary electrons $\mathrm{\langle PE \rangle}$ based on the determination of the mean gain from the fit of our model (solid red line) to the total energy spectrum (black markers). Bottom panel: Relative fraction of events triggering on the SPC channel as a function of the energy. The  error  bars  indicate  the  statistical  (binomial)  uncertainty at the corresponding energy. The red and blue curves show the trigger efficiency curves as derived from the two methods discussed in the core text. These yield a trigger threshold value of $E_{\mathrm{th1}}=0.562~ \mathrm{\langle PE \rangle}$ (resp. $E_{\mathrm{th2}}=0.565~\mathrm{\langle PE \rangle}$) and a standard deviation of $\sigma_{\mathrm{th1}}=0.114~\mathrm{\langle PE \rangle}$ (resp. $\sigma_{\mathrm{th2}}=0.115~\mathrm{\langle PE \rangle}$).
}
\label{Fig:Trig}
\end{center}
\end{figure}

In this section, we present a novel methodology which allows for a precision measurement of the trigger threshold efficiency. We show in Fig.~\ref{Fig:Trig} (top panel) the energy spectrum of laser-induced events (black markers), together with a fit of our model to the data (solid red line) using Eq.~\eqref{eq:pdfFit} and Eq.~\eqref{eq:singlefit}. The contribution to the fit of signal events ($N\geq 1\;e^{-}$) and null events ($N=0\;e^{-}$) are indicated as dashed purple and solid orange lines, respectively. The energy spectrum of events that trigger the SPC channel are indicated as blue markers on the top panel. The bottom panel of Fig.~\ref{Fig:Trig} shows their relative fraction with respect to the total number of events as black markers, together with a fit of our model of the trigger efficiency based on the Gaussian error function (denoted $\mathrm{erf}$) as follows:
\begin{equation}
\epsilon(E_{\mathrm{th}},\sigma_{\mathrm{th}})=\frac{1}{2} \times \left[1 + \mathrm{erf}\left(\frac{E-E_{\mathrm{th}}}{\sqrt{2}\sigma_{\mathrm{th}}}\right) \right]
\label{Eq:erf}
\end{equation}
Because laser-induced events also include null-events (devoid of signal), this method provides a conservative measurement of the trigger threshold efficiency by potentially underestimating it. An alternative two-step approach free of such bias consists in first fitting the total energy spectrum using Eq.~\eqref{eq:singlefit} to determine $\sigma$, $\theta$, $\langle G \rangle$ and $\mu$ and then deriving the expected energy spectrum of signal events ($N\geq 1\;e^{-}$). This model of signal events - corrected for the trigger efficiency using Eq.~\eqref{Eq:erf} - is then fit to the energy spectrum of events triggering on the SPC channel with $E_{\mathrm{th}}$ and $\sigma_{\mathrm{th}}$ as the only free parameters. We show as a solid blue line the result from the fit and the trigger efficiency curve derived from this method on top and bottom panels, respectively. As attested to by the overlap of the two efficiency curves, both methods give essentially identical results when the contribution of the null-events - in the energy range of the determination of the trigger efficiency - is negligible. Although the SPC trigger threshold was voluntarily set in this run to a high value ($E_{\mathrm{th}} \sim 0.5 \;\langle \mathrm{PE} \rangle $) to illustrate the equivalence of the two approaches in such case, this allows us to validate the modeling of the trigger efficiency with Eq.~\eqref{Eq:erf} for its use with the two-step approach in nominal trigger threshold conditions ($E_{\mathrm{th}} \sim 0.2 \;\langle \mathrm{PE} \rangle $).
\subsection{Monitoring of the detector response}
\label{sec:monitoring}
In this section we present the methodology that will be employed by the NEWS-G experiment at SNOLAB to monitor the detector response during dark matter search runs using the UV laser. The approach consists in operating the laser at a high power in order to extract a large number of photo-electrons per event. 
Laser-induced event pulses amplitudes, rise times and delays with respect to PD pulses can be used as probes to monitor the stability of the gain, diffusion and drift time of surface events, respectively. 

\begin{figure}[tpb]
\begin{center}
\includegraphics[width=0.45\textwidth]{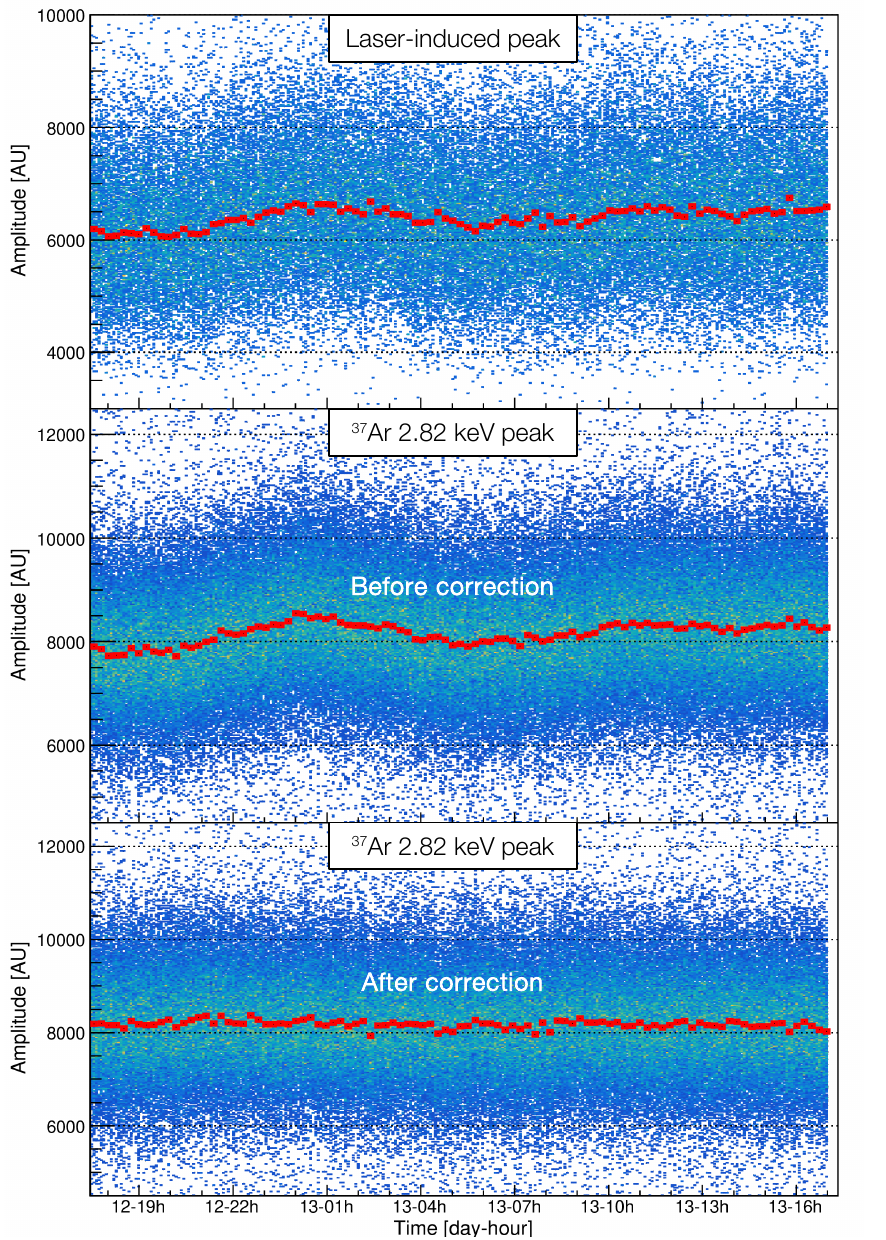}
\caption{Monitoring of the stability of the gain over time using a UV laser. The top panel shows the distribution in SPC pulse amplitude vs. time of laser-induced events corrected for the laser instability using the PD pulses amplitude. The middle and bottom panels show the distribution of ${}^{37}\mathrm{Ar}$ 2822 eV events before (middle panel) and after (bottom panel) correcting for gain variations using the position of laser-induced events. The red markers indicate the center of a Gaussian fitted to amplitude spectra for slices in time of $\sim15\;\mathrm{min}$ width.}
\label{Fig:Stability}
\end{center}
\end{figure}

We show in Fig.~\ref{Fig:Stability} (top panel) the evolution over time of laser-induced events pulse amplitudes (corrected for the laser instability using the PD pulse amplitude) recorded during a $\sim1$~day-long run. The latter was chosen for its distinctive instability of the gain on long-time scales. This arises from significant variations of the room temperature over time (of a few degrees), and therefore of the gas, that we could indirectly measure with a pressure transducer connected to the SPC. We additionally show (middle panel) the distribution of the amplitude of ${}^{37}\mathrm{Ar}$ 2822~eV events recorded during this run. One can see that the position of ${}^{37}\mathrm{Ar}$ and laser-induced events are correlated, indicating that the latter can be used to monitor the stability of the gain. 

To go beyond this solely qualitative assesment, we use the position of the laser-induced peak to apply a time-dependent correction to the amplitude measured for ${}^{37}\mathrm{Ar}$ events (bottom panel). The correction procedure reduces the relative dispersion of the mean 2822~eV peak position by a factor $\sim3$, from $2.6\%$ down to $0.9\%$, demonstrating the ability to monitor gain fluctuations with better than $1\%$ precision. Because the UV-laser is pulsed, it can be used continuously over the whole duration of dark matter search runs without adding any background. The fraction of dead time it induces is the product of the laser pulse rate with the event time window. This corresponds to a 2\% dead time with the operation of the laser at maximum pulse rate (10 Hz) for nominal event time windows of 2 ms. Because space charge effects are known to potentially induce an event rate dependency of the SPC detector response, the laser further presents the advantage - unlike more conventional calibration methods - of monitoring the gain in the same rate conditions as that of the physics run.\\
\begin{figure}[tpb]
\begin{center}
\includegraphics[width=0.44\textwidth]{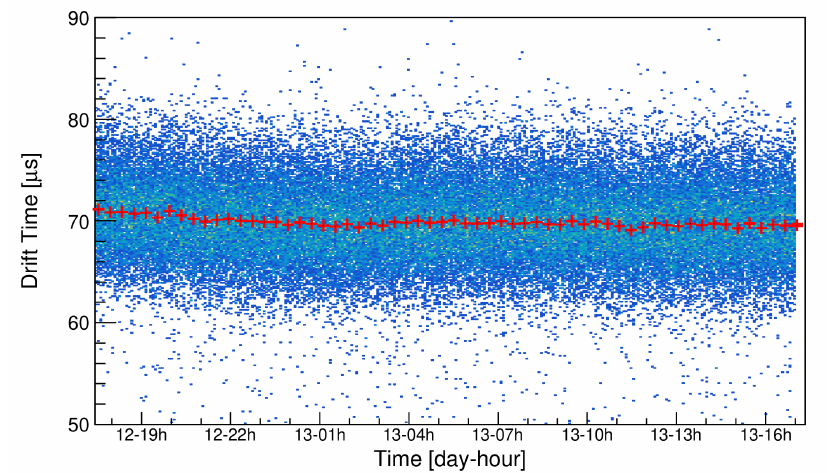}
\caption{Monitoring of the stability of the mean drift time of laser-induced events. the red markers indicate the center of a Gaussian fitted to drift time spectra for slices in time of $\sim20\;\mathrm{min}$ width.}
\label{Fig:DriftStability}
\end{center}
\end{figure}
Laser calibration measurements naturally lend themselves to measurements of the drift time of surface events. The time at which 50\% of the SPC pulse amplitude is reached serves as an estimator of the mean arrival time of the primary electrons. Because the trigger time on the PD channel corresponds to the time at which photo-electrons extracted from the surface start their drift, the mean drift time of surface events is given by the difference between the mean arrival time (SPC channel) and the trigger time (PD channel). 

We show in Fig.~\ref{Fig:DriftStability} the evolution over time of the measured mean drift time for laser-induced events recorded during the same run as shown in Fig.~\ref{Fig:Stability}. It illustrates how the NEWS-G experiment at SNOLAB will be able to continuously monitor the stability of the gas drift properties during WIMP search runs and further demonstrates that drift time measurements of surface events can be achieved with a precision better than $1~\mathrm{\mu s}$. Such measurements will allow us to accurately determine the stability of the efficiency of rise-time based cuts that are aimed at removing surface background events. These will further provide a powerful test of the drift simulation code we rely on for the calculation of the experiment's sensitivity to WIMPs.
\section{Conclusion and Discussion}
In this work, we performed the first characterization of the single electron response of SPCs, using the same laser-based calibration technique that will be used for the upcoming phase of the NEWS-G experiment at SNOLAB. The novel analysis methodology we developed allows for $\mathcal{O}(1\%)$ precision measurements of the mean amplification gain and of the relative gain variance, independently from electron attachment and of laser stability. 
To the best of our knowledge, it is additionally the first time that the SER of a gaseous detector technology is derived from energy spectra containing a significant fraction of events with $\mathrm{N}>1$ electrons. Combining these results with ${}^{37}\mathrm{Ar}$ calibration data, we measured the mean ionization energy $\mathrm{W}=27.6\pm 0.2~\mathrm{eV}$
in $\mathrm{Ne} + \mathrm{CH_4}~(2\%)$ at 1.5~bars for 2822~eV X-rays and demonstrated the feasibility of performing similar precision measurements at sub-keV energies for future gas mixtures to be used for WIMP searches. These will allow us to translate measurements of the quenching factor (ionization yield of nuclear recoils with respect to electron recoils) into absolute measurements of the W-value for nuclear recoils expected from WIMPs. We showed how the laser could be used to accurately measure the stability of the gain and electron drift properties over time as well as to measure experimentally the trigger threshold efficiency down to sub-electron energies. These applications open new avenues for the NEWS-G experiment at SNOLAB which will take advantage of a powerful tool to continously monitor the detector response and signal detection efficiency during WIMP search runs.
\section*{Acknowledgments}
This research was undertaken, in part, thanks to funding from the Canada Excellence Research Chairs Program. The contribution of C.~Garrah to the analysis code is gratefully acknowledged. We also thank A.~Ulrich for useful discussions and crucial insight while conceptualizing the UV-laser calibration system.
\section*{Appendix}
In this section, we discuss two potential sources of systematic uncertainties with respect to the measurement of the Polya parameter $\theta$ and the mean gain $\langle G \rangle$. Although the use of a getter efficiently suppressed electronegative impurities from the gas mixture (see Sec.~\ref{sec:Experimental Set-Up}), one may legitimely wonder how attachment would affect our results. The validity of our model relies in the number $N$ of electrons reaching the avalanche region being Poisson distributed regardless of wether or not a fraction of the $N_{\mathrm{ext}}$ PEs extracted were lost during their drift toward the sensor. In the presence of attachment, and considering each of the $N_{\mathrm{ext}}$ PEs as having the same survival probability $p$, $P(N|N_{\mathrm{ext}},p)$ is governed by the Binomial distribution. Because $N_{\mathrm{ext}}$ is Poisson distributed (see Sec.~\ref{sec:model1}), the probability distribution of the number $N$ of PEs reaching the sensor is given by:
\begin{eqnarray}
P(N|\mu,p)&=&\sum_{n=N_{\mathrm{ext}}}^{\infty} P_{\mathrm{Binomial}}(N|n,p)\cdot P_{\mathrm{Poisson}}(n|\mu) \nonumber \\
 &=&\sum_{n=N_{\mathrm{ext}}}^{\infty} \frac{n!}{N!\,(n-N)!}\cdot p^n(1-p)^{N_{\mathrm{ext}}-n} \cdot \frac{ e^{-\mu}{\mu}^{n} }{n!}  \nonumber \\
 &=&\frac{e^{-\mu} {(p\cdot \mu)}^{N}}{N!} \cdot \sum_{n=N_{\mathrm{ext}}}^{\infty} \frac{{(\mu\cdot(1-p))}^{n-N}}{n-N} \nonumber \\
 &=&\frac{e^{-\mu} {(p\cdot \mu)}^{N}}{N!}\cdot e^{(1-p)\mu} =\frac{e^{-p\mu} {(p\cdot \mu)}^{N}}{N!}
\end{eqnarray}
which is nothing less than $P_{\mathrm{Poisson}}(N|p \cdot \mu)$. Because attachment would still result in $N$ being Poisson distributed, it does not have any impact on the experimental measurement of $\theta$ nor $\langle G \rangle$. However, because attachment would result in an underestimation of the mean number of primary electrons of a mono-energectic line, the associated W-value measurement would be overestimated. 
\begin{figure}[tpb]
\begin{center}
\includegraphics[width=0.49\textwidth]{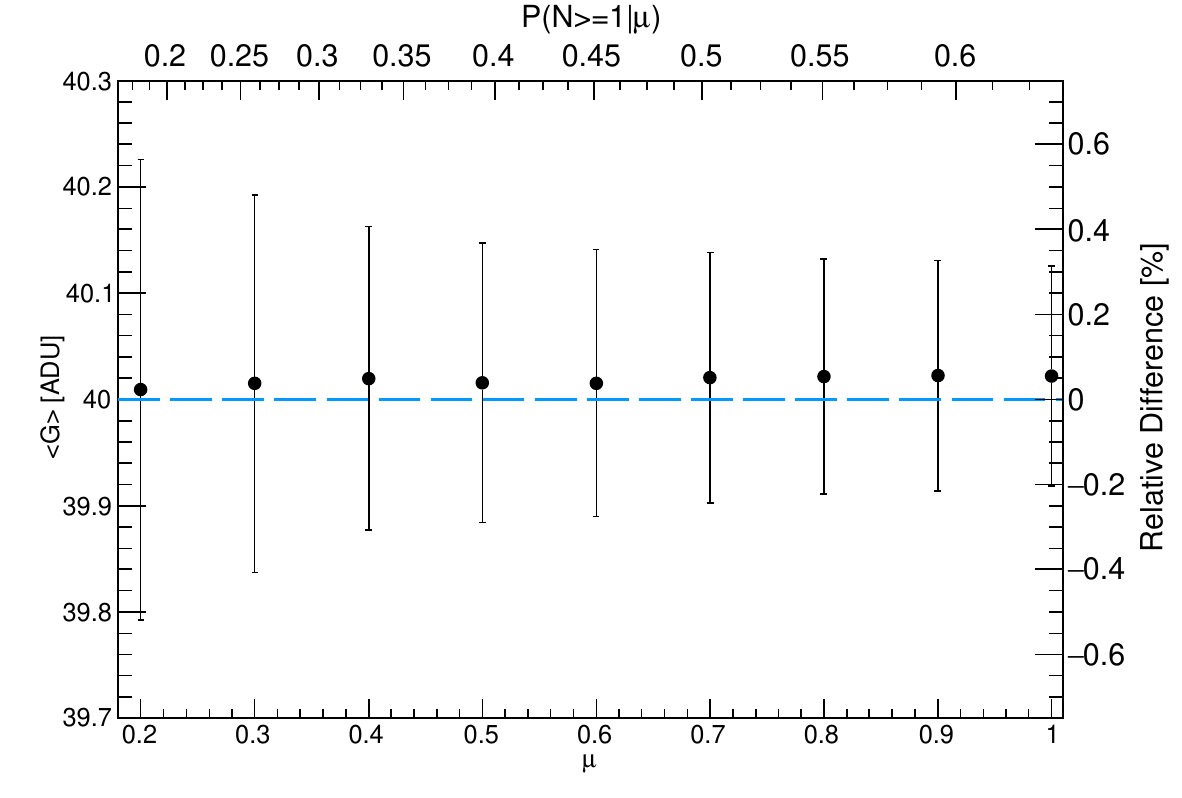}\\
\includegraphics[width=0.49\textwidth]{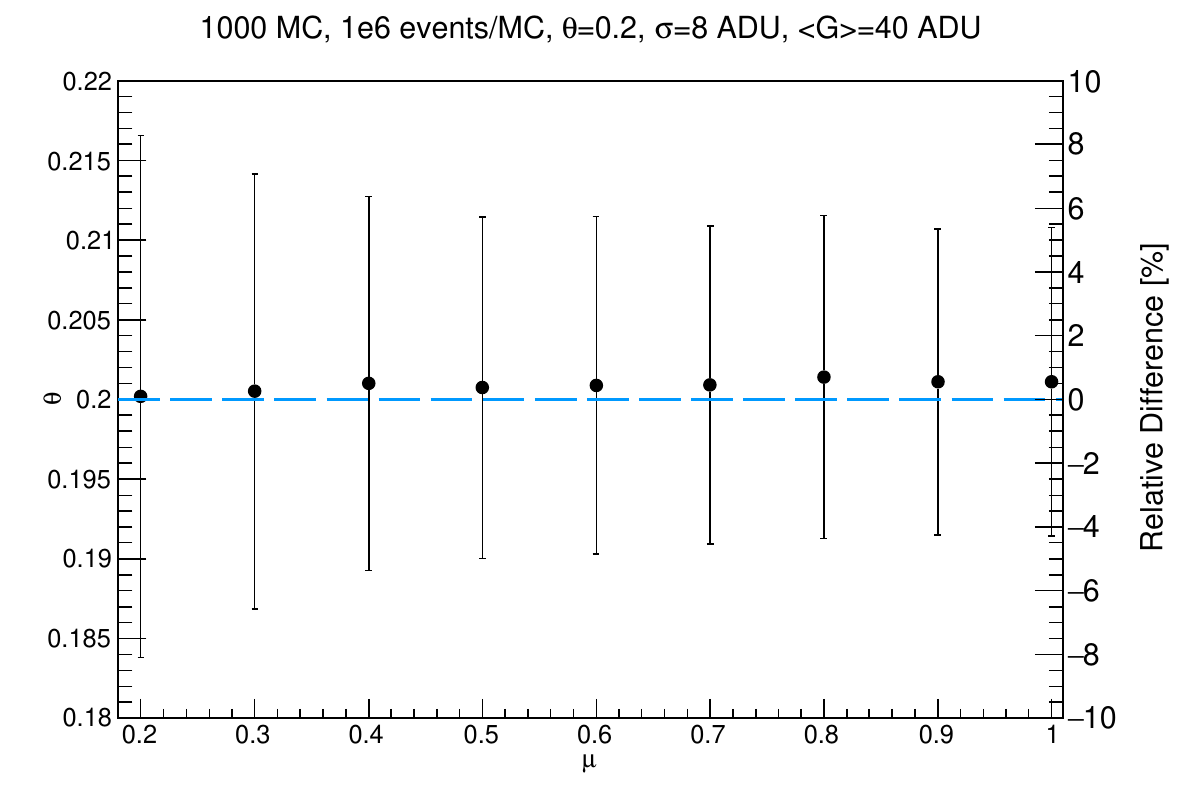}\\
\caption{Effect of the bias induced on the estimators of the mean gain $\langle G \rangle $ (top panel) and of the Polya parameter $\theta$ by $\pm\;5\%$ random fluctuations of the laser power. As explained in more detail in the core text, simulated data sets which are generated by Monte-Carlo account for $\pm5\%$ random fluctuations of $\mu$ on an event-per-event basis. Input values used for the simulations are indicated between the two panels. On each panel, the black markers and the associated error bar are derived from the mean and standard deviation of a Gaussian fit to the distribution of the best fit values of the corresponding parameter of interest. As attested to by their small deviation with respect to the dashed blue line that indicates the input value of $\theta$ and $\langle G \rangle $ in the simulation, the bias is negligible. 
}
\label{Fig:Sys}
\end{center}
\end{figure}

The instability of the laser power is another source of systematic uncertainty associated with our measurements. For a fixed laser power, the number $N$ of electrons reaching the sensor is Poisson distributed with an expectation value of $\mu$. A pulse-to-pulse laser instability will result  in a different expectation value for each of the laser-induced events such that the overall distribution of $N$ is not expected to strictly follow Poisson statistics. Our ability to monitor the laser power on a pulse-to-pulse basis using the PD signals naturaly lends itself to consider performing an unbinned likelihood fit of the SER spectra. Such an analysis approach would account for the different PD signal amplitude of each individual event and would simply consist in replacing the free parameter $\mu$ in the fit by its linear conversion factor with the PD amplitude (i.e. the slope of the line in top left panel of Fig.~\ref{Fig:Fit2}). However, because the computing time of an unbinned likelihood fit is directly proportional to the number of events, the high statistics of laser-induced events makes this approach impractical. Instead, our method consists in dividing each run into sub-data sets based on the PD amplitude of the events. Because the tolerance of the cuts in PD amplitude is set to $\pm5\%$ to provide us with sufficient statistics, we investigated the bias a fluctuating value of $\mu$ may induce on our estimators of $\langle G \rangle$ and $\theta$. To achieve this, $10^3$ SER energy spectra consisting of $10^6$ events per data set were simulated by Monte-Carlo for different values of $\mu$. For each event, the number $N$ of electrons is drawn from the Poisson distribution with an expectation value randomly drawn between $0.95\cdot\mu$ and $1.05\cdot\mu$. The energy associated with each event is drawn from the $N^{\mathrm{th}}$ convolution of the Polya distribution smeared with a Gaussian to simulate the effect of baseline noise. Then, the energy spectrum is fitted using Eq.~\eqref{eq:pdfFit} and Eq.~\eqref{eq:singlefit}. For each value of $\mu$, we fit a Gaussian to the distribution of the $10^3$ best fit values of our parameters of interest. The mean values are reported as black markers in Fig.~\ref{Fig:Sys} as a function of $\mu$, both for $\langle G \rangle$ (top panel) and $\theta$ (bottom panel). The dashed blue line in each panel indicates - for the corresponding parameter of interest - the value that was used as an input to simulate the data sets. Interestingly, the bias induced by $\pm5\%$ fluctuations of $\mu$ on the estimator of $\langle G \rangle$ is $<0.1\%$. The impact on the estimator of $\theta$ is also extremely small, with $<1\%$ deviation from the value of $\theta=0.2$ used as an input. We give hereafter elements of a qualitative explanation for the smallness of the bias induced by fluctuations of $\mu$ using the following formula:
\begin{eqnarray}
\lim_{\varepsilon \to 0}\frac{P(N|(1-\epsilon)\mu)+P(N|(1+\epsilon)\mu)}{2} \quad \quad \nonumber \\
=P(N|\mu)\times \left(1+\frac{\epsilon^2}{2}(N^2-3N+1)+\mathcal{O}(\epsilon^3)\right)
\label{Eq:explanation}
\end{eqnarray}\\
Eq.~\eqref{Eq:explanation} illustrates how the departure from the Poisson distribution $P(N|\mu)$ of the overall distribution of $N$ is of a second order with respect to the fluctuation in $\mu$. As an example, a value of $\epsilon=5\%$ in Eq.~\eqref{Eq:explanation} yields a $\sim0.1\%$ deviation from the Poisson distribution. Fig.~\ref{Fig:Sys} illustrates another interesting effect, which while not related to the bias of our estimators is worth mentionning. One can see that the errors bars, which give an indication of the precision with which parameters of interest can be measured, get smaller with increasing values of $\mu$. These typically improve with the square root of the increase in the number of signal events ($P(N\geq 1 |\mu)$ depicted on top axis) due to null-events not bringing any other information than the baseline resolution. This illustrates one of the numerous advantages confered by the ability to fit SER spectra containing a significant proportion of events with $N>1$ electrons. 
\bibliography{BibliographyLaserPaper}
\end{document}